\documentclass[twocolumn]{aastex63}

\NewPageAfterKeywords
\usepackage{needspace}

\turnoffedit

\received{\_\_\_\_\_\_\_\_\_\_}
\revised{\_\_\_\_\_\_\_\_\_\_}
\accepted{\_\_\_\_\_\_\_\_\_\_}
\submitjournal{ApJ}

\shorttitle{WB-57 Observations of the 2017 Total Eclipse}
\shortauthors{Caspi et al.}

\begin{document}

\title{A new facility for airborne solar astronomy:\\
NASA's WB-57 at the 2017 total solar eclipse}


\correspondingauthor{Amir Caspi}
\email{amir@boulder.swri.edu}


\author[0000-0001-8702-8273]{Amir Caspi}%
\affil{Southwest Research Institute,
Boulder, CO 80302 USA}

\author[0000-0002-0494-2025]{Daniel B. Seaton}%
\affiliation{Cooperative Institute for Research in Environmental Sciences,
University of Colorado,
Boulder, Colorado, 80305 USA}
\affiliation{National Centers for Environmental Information,
National Oceanic and Atmospheric Administration,
Boulder, Colorado, 80305 USA}

\author[0000-0002-1939-6813]{Constantine C. C. Tsang}%
\affil{Southwest Research Institute,
Boulder, CO 80302 USA}

\author[0000-0002-7164-2786]{Craig E. DeForest}%
\affil{Southwest Research Institute,
Boulder, CO 80302 USA}

\author[0000-0001-5681-9689]{Paul Bryans}%
\affil{High Altitude Observatory,
National Center for Atmospheric Research,
Boulder, CO 80301 USA}

\author[0000-0001-7416-2895]{Edward E. DeLuca}%
\affil{Center for Astrophysics {\textbar} Harvard \& Smithsonian,
Cambridge, MA 02138 USA}

\author[0000-0001-7399-3013]{Steven Tomczyk}
\affil{High Altitude Observatory,
National Center for Atmospheric Research,
Boulder, CO 80301 USA}

\author[0000-0002-9959-6048]{Joan T. Burkepile}%
\affil{High Altitude Observatory,
National Center for Atmospheric Research,
Boulder, CO 80301 USA}

\author{Thomas ``Tony'' Casey}%
\affil{Southern Research,
Houston, TX 77034 USA}

\author{John Collier}
\affil{Southern Research,
Birmingham, AL 35211 USA}

\author{Donald ``DD'' Darrow}%
\affil{Southern Research,
Houston, TX 77034 USA}

\author{Dominic Del Rosso}
\affil{NASA Johnson Space Center,
Houston, TX 77034 USA}

\author[0000-0003-4782-1503]{Daniel D. Durda}%
\affil{Southwest Research Institute,
Boulder, CO 80302 USA}

\author[0000-0001-9745-0400]{Peter T. Gallagher}%
\affiliation{School of Physics, Trinity College Dublin,
Dublin 2, Ireland}
\affiliation{School of Cosmic Physics, Dublin Institute for Advanced Studies,
Dublin 2, Ireland}


\author[0000-0001-9638-3082]{Leon Golub}%
\affil{Center for Astrophysics {\textbar} Harvard \& Smithsonian,
Cambridge, MA 02138 USA}

\author{Matthew Jacyna}%
\affil{Viasat, Inc.,
Carlsbad, CA 92009 USA}

\author{David ``DJ'' Johnson}
\affil{NASA Johnson Space Center,
Houston, TX 77034 USA}

\author[0000-0001-5174-0568]{Philip G. Judge}%
\affil{High Altitude Observatory,
National Center for Atmospheric Research,
Boulder, CO 80301 USA}

\author{Cary ``Diddle'' Klemm}
\affil{NASA Johnson Space Center,
Houston, TX 77034 USA}

\author[0000-0003-0818-680X]{Glenn T. Laurent}%
\affil{Southwest Research Institute,
Boulder, CO 80302 USA}

\author{Johanna Lewis}
\affil{Southern Research,
Birmingham, AL 35211 USA}

\author{Charles J. Mallini}
\affil{NASA Johnson Space Center,
Houston, TX 77034 USA}

\author{Thomas ``Duster'' Parent}
\affil{NASA Johnson Space Center,
Houston, TX 77034 USA}

\author{Timothy Propp}
\affil{NASA Johnson Space Center,
Houston, TX 77034 USA}

\author[0000-0002-5358-392X]{Andrew J. Steffl}%
\affil{Southwest Research Institute,
Boulder, CO 80302 USA}

\author{Jeff Warner}%
\affil{Viasat, Inc.,
Carlsbad, CA 92009 USA}

\author[0000-0002-0631-2393]{Matthew J. West}%
\affil{Solar-Terrestrial Centre of Excellence -- SIDC, Royal Observatory of Belgium,
1180 Brussels, Belgium}

\author{John Wiseman}
\affil{Southern Research,
Birmingham, AL 35211 USA}

\author{Mallory Yates}
\affil{NASA Johnson Space Center,
Houston, TX 77034 USA}

\author[0000-0002-2542-9810]{Andrei N. Zhukov}%
\affiliation{Solar-Terrestrial Centre of Excellence -- SIDC, Royal Observatory of Belgium,
1180 Brussels, Belgium}
\affiliation{Skobeltsyn Institute of Nuclear Physics, Moscow State University,
119991 Moscow, Russia}

\author{the NASA WB-57 2017 Eclipse Observing Team}
\noaffiliation

\begin{abstract}
NASA's WB-57 High Altitude Research Program provides a deployable, mobile, stratospheric platform for scientific research. Airborne platforms are of particular value for making coronal observations during total solar eclipses because of their ability both to follow the Moon's shadow and to get above most of the atmospheric airmass that can interfere with astronomical observations. We used the 2017~Aug~21 eclipse as a pathfinding mission for high-altitude airborne solar astronomy, using the existing high-speed visible-light and near-/mid-wave infrared imaging suite mounted in the WB-57 nose cone. In this paper, we describe the aircraft, the instrument, and the 2017 mission; operations and data acquisition; and preliminary analysis of data quality from the existing instrument suite. We describe benefits and technical limitations of this platform for solar and other astronomical observations. We present a preliminary analysis of the visible-light data quality and discuss the limiting factors that must be overcome with future instrumentation. We conclude with a discussion of lessons learned from this pathfinding mission and prospects for future research at upcoming eclipses, as well as an evaluation of the capabilities of the WB-57 platform for future solar astronomy and general astronomical observation.

\end{abstract}

\keywords{Astronomical instrumentation (799), Solar instruments (1499), Solar eclipses (1489), Total eclipses (1704), Solar corona (1483), Solar coronal heating (1989), Solar coronal plumes (2039), Solar coronal streamers (1486), Solar coronal waves (1995), The Sun (1693)}

\section{Introduction}\label{sec:intro}
Although high-altitude and space-based instruments have observed the solar corona since the late 1950s \citep{1957Natur.179..861C, 1959JGR....64..697P}, the launch of the \textit{Solar and Heliospheric Observatory} (\textit{SOHO}) with its Extreme-Ultraviolet Imaging Telescope \citep[EIT;][]{Delaboud95} and Large Angle Spectroscopic Coronagraph \citep[LASCO;][]{1995SoPh..162..357B} marked a turning point in the way we study the corona. Since that time, the vast majority of solar coronal research has used space-based observatories. However, for certain applications of coronal physics, studies of the corona during total solar eclipses nonetheless offer considerable advantages over observations from space. In particular, constraints of on-board storage and telemetry put limits on the rate at which data can be acquired from space, foreclosing many science objectives that require extremely high observing cadences. Likewise, technical constraints on the optical and electronic systems used for space-based observations of the corona, both by ultraviolet and X-ray telescopes and, in visible light, by coronagraphs, impose their own limitations on the detectability of some phenomena --- particularly faint or highly dynamic features.

Total solar eclipses, therefore, provide a unique opportunity to explore a variety of problems in coronal physics that cannot be solved with space-based observations. Of particular historical importance are programs of eclipse observations oriented towards explaining the mechanism that heats the corona to a few million~K \citep{2009Natur.459..789P}. Eclipse studies, because they can yield observations with very high spatial and temporal resolution, have the potential to distinguish between mechanisms described by competing theories of coronal heating, primarily the dissipation of magnetic energy from various oscillations in the corona and direct release of magnetic energy via reconnection \citep{2003A&ARv..12....1W, Klimchuk2006}.



A challenge of observing during eclipses using traditional ground-based methods, however, is that the spatial and temporal scales of these dynamic processes are similar to the scales on which Earth atmospheric turbulence acts to distort observations (``seeing''). Additionally, certain wavelength bands (e.g., in the infrared) are strongly contaminated by atmospheric absorption and/or emission, significantly compromising or prohibiting useful observations in these bands from the ground. Furthermore, eclipses provide only a brief window -- a few minutes at most -- to observe the corona. Using mobile, airborne platforms to observe total solar eclipses overcomes all of these limitations by operating above most of the atmospheric airmass and water vapor content, and by enabling a greater duration within the moving shadow.

A notable early attempt to observe an eclipse from an airborne platform was from the dirigible U.S.S. Los Angeles near Montauk Point on Long Island, NY during the total solar eclipse of 1925 January 24 \citep{1933PUSNO..13B..73.}, which observed from an altitude of 4,500 feet ($\sim$1370 m) using still and motion picture cameras and a spectrograph. The initial campaign proposal called for a platform to be constructed on the top of the airship to hold the eclipse-observing apparatus and personnel, but this was deemed infeasible due in large part to technical limitations of such construction and the prevailing weather conditions during deep winter in New York. Instead, observations were made through windows in the passenger cabin. The rationale for using an airship for this observation was to improve the odds of a successful observation in the event of low-level cloudy conditions (although ground conditions turned out to be favorable). Their observations were indeed successful, but due to the motion of the airship caused by wind, the spectrographic data were ``of no value for accurate wave-length determinations.''\footnote{\rightskip=10mm Capt. Littell's account of this expedition is highly entertaining and informative. We strongly recommend it to the reader. It is interesting to note that some of the same difficulties they encountered in making their airborne measurements are still of concern in our observations made nearly 100 years later.}

Perhaps the most famous airborne eclipse experiment is the flight of the Concorde at the 1973 eclipse over North Africa, whose supersonic flight along the eclipse track afforded a 74-minute window of totality \citep{1973Natur.246...72B, 1975LAstr..89..149K} yielding both visible and infrared (IR) observations of the corona. However, their measurements were significantly limited by seeing effects due to turbulence in the sonic shock around the aircraft. \citet{Kuhn1999} observed static images of the near-IR (1--4~$\mu$m) corona in various spectral lines from a C130 aircraft during the 1998 eclipse over the Pacific Ocean. Other eclipses have been observed from airplanes for purely logistical reasons, due to the inaccessibility of the terrain over which the eclipse passed; these include the 2003 total solar eclipse over Antarctica and the 2008 solar eclipse that traversed a wide range of longitudes near the North Pole \citep{1674-4527-9-6-001}. A handful of eclipses have also passed directly over mountaintop astronomical observatories or highly accessible high-altitude terrain, such as the 1991 solar eclipse over Mauna Kea, Hawai'i which produced, for example, IR observations that could not have been made from lower altitudes due to atmospheric absorption and background emission of IR radiation \citep{1992Natur.355..707H}.

The 2017~August~21 total solar eclipse crossed the entire continental United States, the first to do so since 1918, and was the best observed eclipse to date \citep{Miller2018}. This eclipse presented unique opportunities not only for public engagement and community science \citep[e.g.,][]{2015SoPh..290.2381K, Penn2017}, but also for professional research. For example, \citet{2018ApJ...856L..29S} used a new instrument, the Airborne Infrared Spectrometer (AIR-Spec), flying on the National Center for Atmospheric Research's High-performance Instrumented Airborne Platform for Environmental Research (HIAPER) Gulfstream-V aircraft, to obtain IR spectra of the corona between 1.4 and 4~$\mu$m with four slit positions at different locations near the solar limb.
The HIAPER has a $\sim$10-hour range at altitudes up to $\sim$45~kft and can accommodate fairly sizable instrument packages within the passenger cabin but, as a result, the fields-of-view are restricted through specific quartz windows that must be configured before flight, and coarse pointing is tied to the aircraft heading and orientation.

NASA's WB-57F high-altitude research aircraft offers an agile new facility for airborne astronomy. With an integrated, nose-mounted, stabilized pointing platform, they offer unrestricted visibility to celestial targets anywhere in the forward hemisphere of the aircraft, and some of the rear, with broad flexibility in aircraft heading and orientation. With a higher service ceiling of up to $\sim$65~kft and a range of $\sim$6~hours, multiple available aircraft enable simultaneous multi-point, multi-instrumented, and/or multi-target observing campaigns.

In this paper, we describe our pathfinding mission using two NASA WB-57F aircraft in a coordinated campaign to observe the 2017 eclipse. Our primary scientific motivation for developing the WB-57 as a platform for airborne eclipse studies is to enable high-resolution, high-cadence observations of the visible-light and infrared corona, uncontaminated by atmospheric seeing effects. Such observations are critical for studies of the rapid dynamics associated with the high-energy processes postulated to be responsible for coronal heating. For this first mission, to accommodate a stringently limited budget and schedule, we used the existing imaging instrument suite already in place on the nose-mounted pointing platform, observing at an altitude of approximately 52.9~kft ($\sim$16.1~km) in both visible light
and in near/mid-wave (3--5~$\mu$m) IR. We used these measurements to characterize the facility performance for astronomical observations and specifically for eclipse studies, including preliminary scientific analysis to further quantify the capabilities of the existing instrument suite and guide future instrument development. Flying the two airplanes in tandem, separated by approximately 100~km, allowed us to test aircraft coordination and provided approximately 7.5 continuous minutes of totality. We also made opportunistic observations of Mercury during the pre- and post-totality partial phases of the eclipse, and calibration observations of stars and Venus on the inbound and outbound ferry segments. Here, we describe this pathfinding mission, our prototype observations, and analysis of platform performance, including data quality and limitations of the existing instrumentation. We discuss a preliminary scientific analysis to place the prototype data in context of our motivating question of coronal heating, and conclude with suggestions and outlook for future missions and instrument upgrades.

\needspace{3\baselineskip}
\section{Platform and Instrumentation} \label{sec:inst}

\begin{figure*}[!ht]
    \centering
    \includegraphics[height=0.16\textheight]{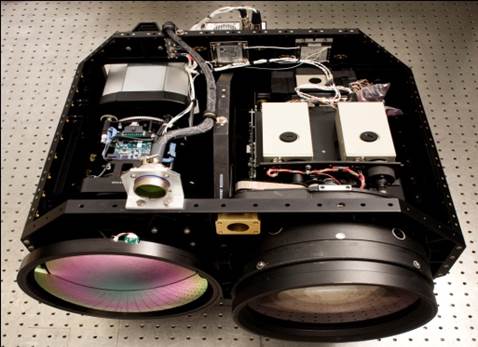}
    \includegraphics[height=0.16\textheight]{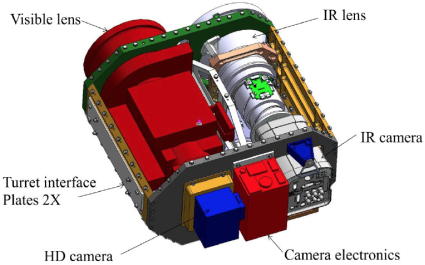}
    \includegraphics[height=0.16\textheight]{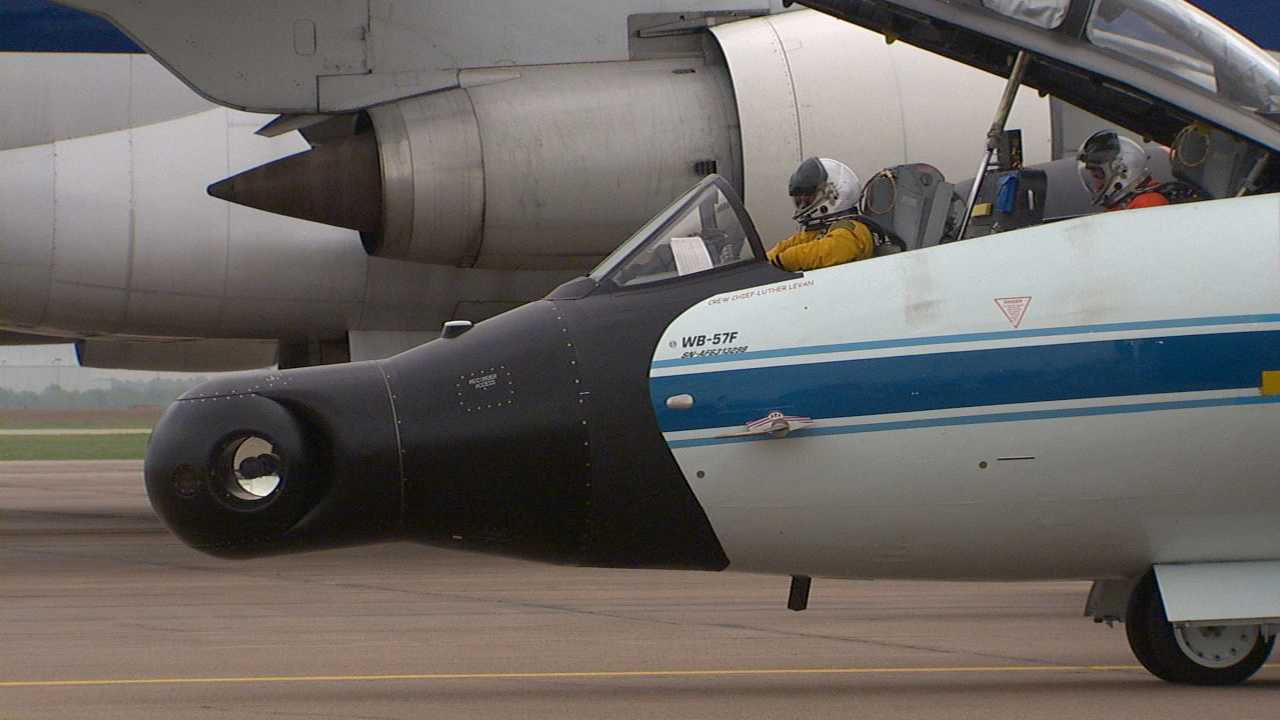}
    \caption{Image (left) and schematic (center) of the DyNAMITE camera system, and the AIRS mounting position in the nosecone of the WB-57 (right).}
    \label{fig:AIRS_overview}
\end{figure*}

NASA's WB-57 High Altitude Research Program\footnote{\url{https://airbornescience.nasa.gov/aircraft/WB-57_-_JSC}} provides a stratospheric (up to 65~kft / 19.8~km) platform for scientific research. We leverage the Airborne Imaging and Recording System (AIRS), a two-axis stabilized pointing platform that mounts in the nose cone. Two identical AIRS units were developed and are maintained by Southern Research (SR) under contract from NASA. AIRS provides visibility for targets anywhere in the forward hemisphere of the aircraft, and some of the rear, enabling continuous target tracking largely independent of aircraft heading and motion. It records pointing information and aircraft position and orientation at 100 Hz with GPS time-tagging, which, in principle, can be used to determine camera position and orientation on the sky.

For this prototype mission, we used existing imaging instrumentation for the AIRS platform. The Day/Night Airborne Motion Imagery for Terrestrial Environments (DyNAMITE) instrument comprises an optical bench with a $1920\times1080$-pixel high definition (HD) broadcast television camera coupled with a zoom lens for visible-light observations, and a scientific-grade infrared camera with actively-cooled InSb detector ($1344\times784$~pixels) coupled with a filtered lens for mid-wave infrared (MWIR; 3--5~$\mu$m) observations\footnote{For complete DyNAMITE instrument specifications, contact the WB-57 Program Office.}. The visible and MWIR apertures are each $\sim$220~mm in diameter, and we operated both cameras at 30~Hz and exposure times of $\sim$30~ms. Figure~\ref{fig:AIRS_overview} shows an overview of the AIRS/DyNAMITE instrument and its location on the WB-57.

The visible-light system's zoom lens provides a selectable field of view (FOV); 
for our eclipse observations, we operated with a 1.6$^{\circ} \times 0.9^{\circ}$ FOV ($\sim$3{\arcsec}/pixel scale) that, with the Sun offset in the image frame, allowed imaging to heights of about 4~R$_\mathrm{Sun}$ towards the west and 2~R$_\mathrm{Sun}$ to the east. The IR lens was operated with a smaller FOV of 1.2$^{\circ} \times 0.7^{\circ}$ (also with $\sim$3{\arcsec}/pixel scale) and images to correspondingly lower heights above the limb. The FOVs were oriented such that the long axes were approximately aligned with the solar equator and the short axes with the poles.

For this flight, the visible-light channel was outfitted with a green ($533.9\pm4.75$~nm) filter. Because the DyNAMITE camera uses a three-channel RGB color system, the primary benefit of such a monochromatic filter is to isolate all of the measured signal into a single color channel. This greatly simplifies the resultant data analysis since all of the photometric information is thus encoded purely within the signal intensity (i.e., the luma channel of the encoded output, which has twice the resolution of the chroma channel; see Section~\ref{sec:data}).

The specific filter passband was chosen to capitalize on a secondary benefit, namely, to leverage a unique spectral feature of the solar corona: the ``coronium'' Fe~\textsc{xiv} 530.3~nm line emitted by hot ($\sim$2~MK) plasma confined to coronal loops. This line is intrinsically narrow \citep[$\sim$1~{\AA}, e.g.,][]{Mierla2008}, so traditionally, many eclipse studies utilize narrowband ($\lesssim$0.5~nm) green filters to strongly isolate the line. In combination with off-band observations, it is possible to quantify the continuum contribution and thus to derive the specific line intensity to enable temperature-dependent diagnostics. However, within the limitations of the existing AIRS/DyNAMITE system that we used due to logistical constraints discussed above, the tradeoffs required to optimize the wide FOV and sensitivity to dim features while minimizing saturation necessitated the use of a low $f$-number ($f$/1.6). In turn, this required a moderately broad ($\sim$9.5~nm FWHM) passband to ensure a relatively uniform response to this wide-angle beam across the entire filter. Additionally, the resource constraints did not permit the installation of a filter wheel that would enable off-band observations. The critical consequence of these circumstances is that our observations of this event cannot be used for temperature diagnostics, but they nonetheless provide information on coronal structure and dynamics and, crucially, allow us to evaluate the platform and instrument performance in general. Moreover, the presence of the green line within our passband nonetheless improves the contrast of hot coronal loops against the diffuse, continuum-dominated, Thomson-scattered K- and dust-scattered F-corona background compared to observations made without a filter. 

\begin{figure}
    \centering
    \includegraphics[width=1.0\columnwidth]{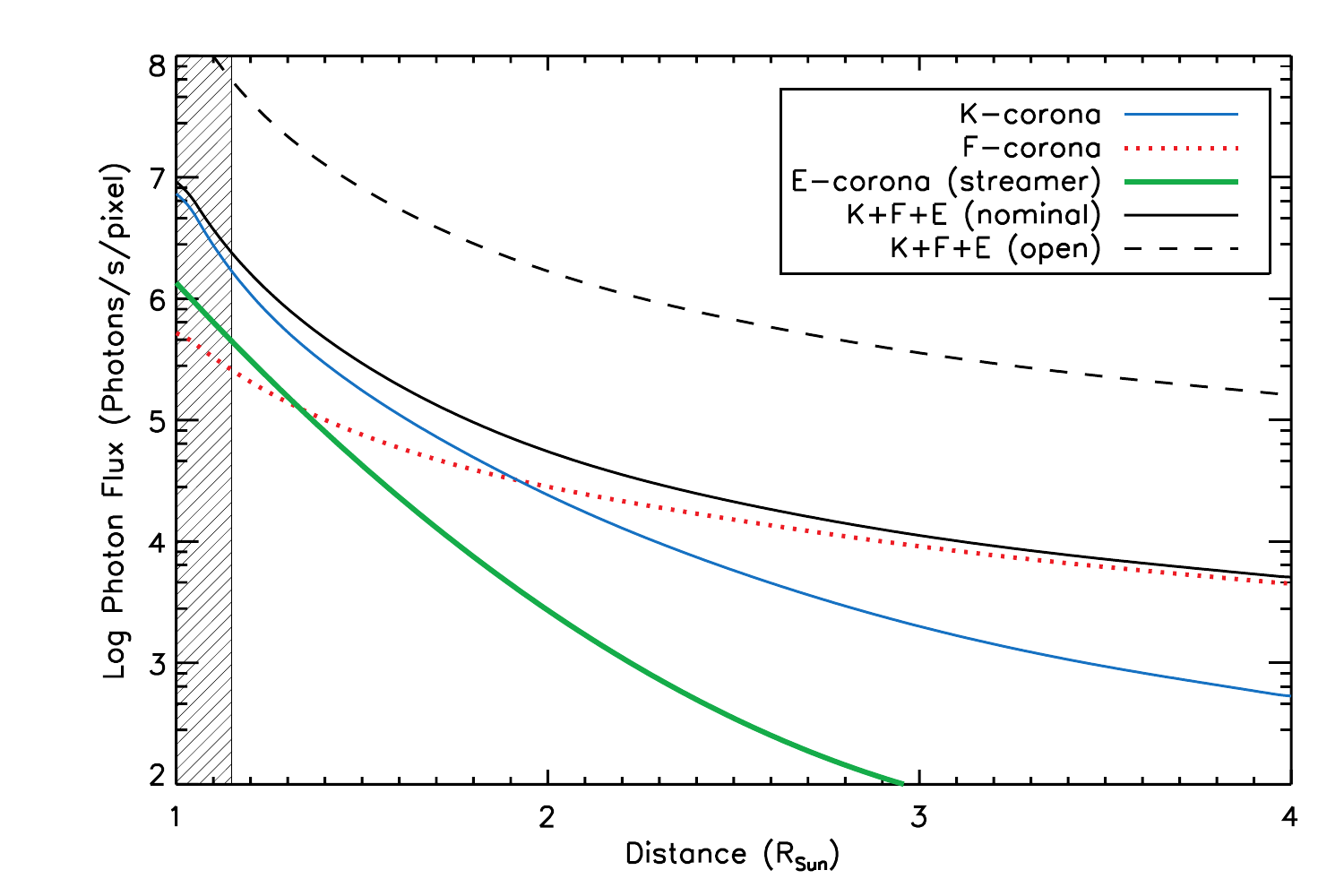}
    \caption{Pre-flight predicted brightnesses of the K-, F-, and green-line E-coronae, and total, in an equatorial streamer as a function of height in the DyNAMITE field of view, for a moderately broad filter with 533.9~nm central wavelength and 9.5~nm FWHM bandwidth, as flown. The total K+F+E signal expected from a wide ``open'' (200~nm) filter (\textit{dashed}) is shown for comparison, highlighting the contrast increase from the green filter. The K- and F-corona values are adapted from \citet{1953sun..book..207V} and green-line values from  \citet{1997ASIC..494..159K}, and are folded through the DyNAMITE optical parameters and the filter bandpasses as stated. The hashing indicates the approximate region of inner-corona saturation in the actual observations reported in Section~\ref{sec:data}.}
    \label{fig:bandpass}
\end{figure}

Figure~\ref{fig:bandpass} shows our pre-flight estimate of the contributions of the K-, F-, and Fe~\textsc{xiv} green-line E-coronae to the total anticipated signal as a function of distance from solar center for our optical setup. The K- and F-corona intensities were adapted from measurements by \citet{1953sun..book..207V}, while the green-line intensity was derived from measurememts by \citet{1997ASIC..494..159K}. These estimates show that the green line should contribute a maximum of $\sim$20\% of the total signal at distances of 1.1--1.3~R$_\mathrm{Sun}$ and remains greater than $\sim$5\% out to 2~R$_\mathrm{Sun}$. This fractional contribution is at least an order of magnitude greater than if we integrated over the entire green-channel bandwidth (estimated at 200~nm) without a filter. It is important to note that the E-corona is highly structured and therefore these estimates should be taken as averages over a variety of conditions, while specific cross-sections through the corona could vary significantly. Crucially, tuning the filter in this way potentially achieves some additional benefit through contrast enhancement and in no way degrades the observations compared to other choices.

\needspace{3\baselineskip}
\section{Mission Design and Operations} \label{sec:mission}

AIRS/DyNAMITE is operated during flight by a Sensor Equipment Operator (SEO) sitting in the aft seat of the aircraft. The pilot, in the forward seat, is responsible for flight operations while the SEO manages the scientific operations of the mission. A high-speed satellite feed from the aircraft enables scientists and engineers on the ground to share the SEO's ``heads-up display'' of live DyNAMITE data, and to communicate with the SEO to provide feedback on instrument settings to optimize the quality of the observations. (For this mission, the live satellite feed also provided the opportunity to stream real-time eclipse imagery to NASA's public outreach television and internet coverage.)

AIRS includes a database and conversion algorithm to determine and maintain the appropriate elevation and azimuth for stable pointing to particular astronomical sources depending on the plane's instantaneous location and attitude, including compensation for maneuvers, so it is not necessary for the SEO to manually maintain pointing throughout the observing period. We used the pre-eclipse prediction of coronal structure from Predictive Science, Inc. \citep{2018NatAs...2..913M} to determine our desired offset from Sun-center pointing, to provide observations higher above the limb with the highest-priority target for potential high-speed dynamics.

The SEO must manually set various camera parameters such as exposure time and focus, some of which may need to be adjusted during the observing period to optimize performance. We prepared a detailed, step-by-step observing procedure that outlined each observing target and the associated camera settings that we anticipated from first-order estimates prior to flight. Some parameters, such as focus, could not be estimated \textit{a priori} because they depend on actual, in-flight circumstances and thus had to be adjusted on-the-fly by the SEO. We note that DyNAMITE does not currently include quantitative focus feedback, and the visible-light camera also does not report image dynamic range statistics, so optimizing focus and exposure time relied on the SEO's visual estimation.

\begin{figure}[!b]
    \centering
    \includegraphics[width=1.0\columnwidth]{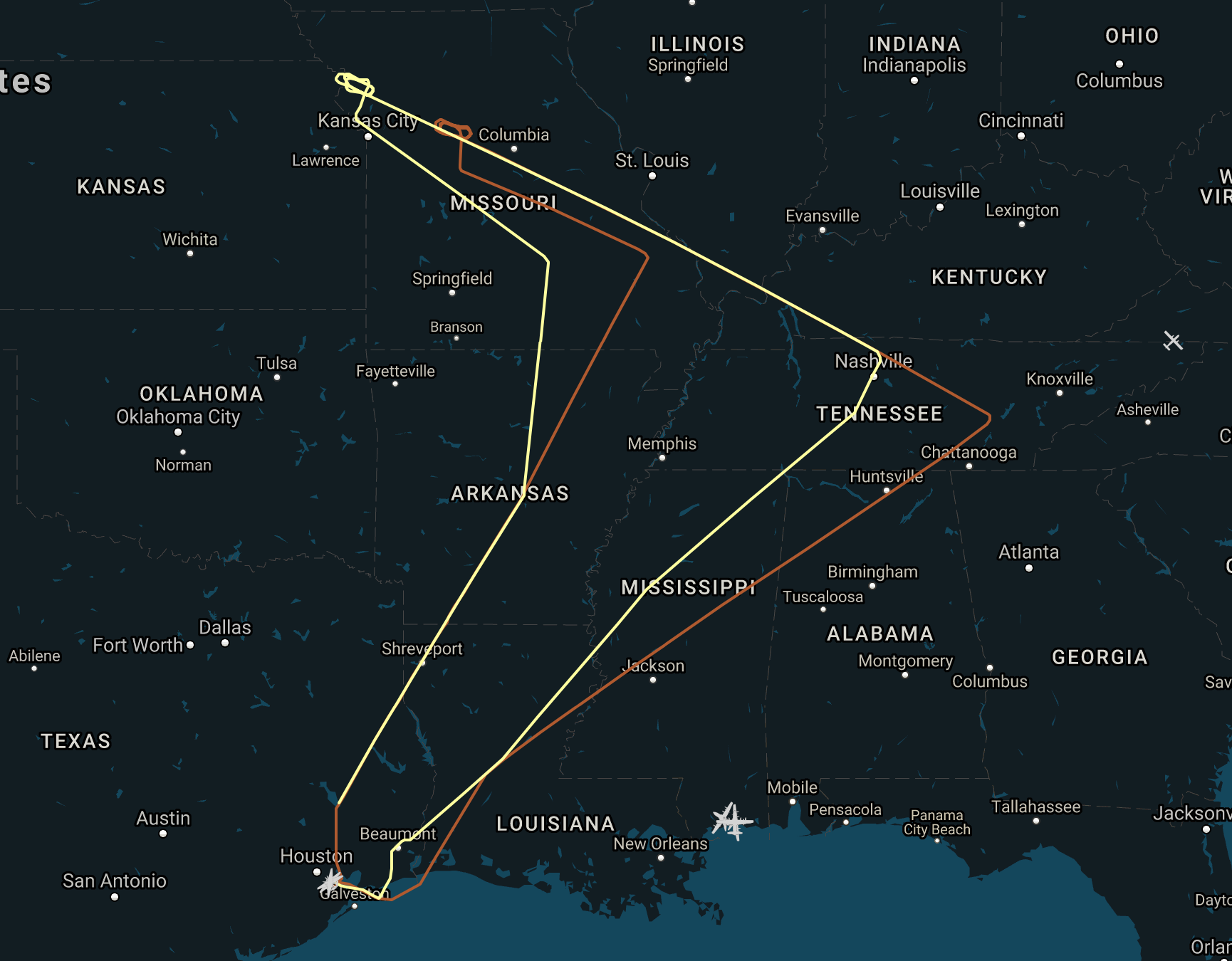}
    \caption{Ground tracks for the two WB-57 aircraft as flown during the mission. Calibration images were obtained during the outbound and return segments. Timing control loops (\textit{top left}) were used to stage the airplanes prior to the eclipse observation segment that proceeded west to east. Totality occurred west-southwest of St. Louis for the trailing airplane (\textit{yellow}) and over Carbondale, roughly due west from Evansville, for the leading plane (\textit{brown}). (Track information from \url{https://airbornescience.nasa.gov/tracker/})}
    \label{fig:flight_track}
 \end{figure}

Our eclipse flight plan called for two aircraft to operate in tandem, traveling at 750~kph ($\sim$400~knots) along the eclipse path at an altitude of 15.2~km ($\sim$50~kft), with a maximum separation of 105~km. (The actual altitude during observation was $\sim$16.1~km [$\sim$52.9~kft].) This flight configuration provided $\sim$50\% greater observing time per airplane compared to a stationary observer ($\sim$240~s vs. $\sim$160~s near eclipse greatest duration) and ensured sufficient overlap in the observations from the respective airplanes to unify their two independent data sets into a single continuous observation of totality.

Figure~\ref{fig:flight_track} shows the ground tracks for the aircraft during the eclipse. Calibration images were taken during the outbound and return legs of the trip. After arriving at their respective starting locations on the eclipse path, the airplanes entered a staging pattern for accurate timing control before heading down the track at the correct time. The two airplanes each observed Mercury during the partial phases of the eclipse, before and after imaging the corona during totality. The trailing aircraft observed totality first, from 18:16:07~UT to 18:20:09~UT; the lead aircraft then observed totality from 18:19:47~UT to 18:23:44~UT. This resulted in 22~s of totality overlap, and a total duration of 457~s of continuous coronal observations during totality.

\needspace{3\baselineskip}
\section{Data and Calibration} \label{sec:data}

\begin{figure*}[!b]
    \centering
    \includegraphics[width=0.95\textwidth]{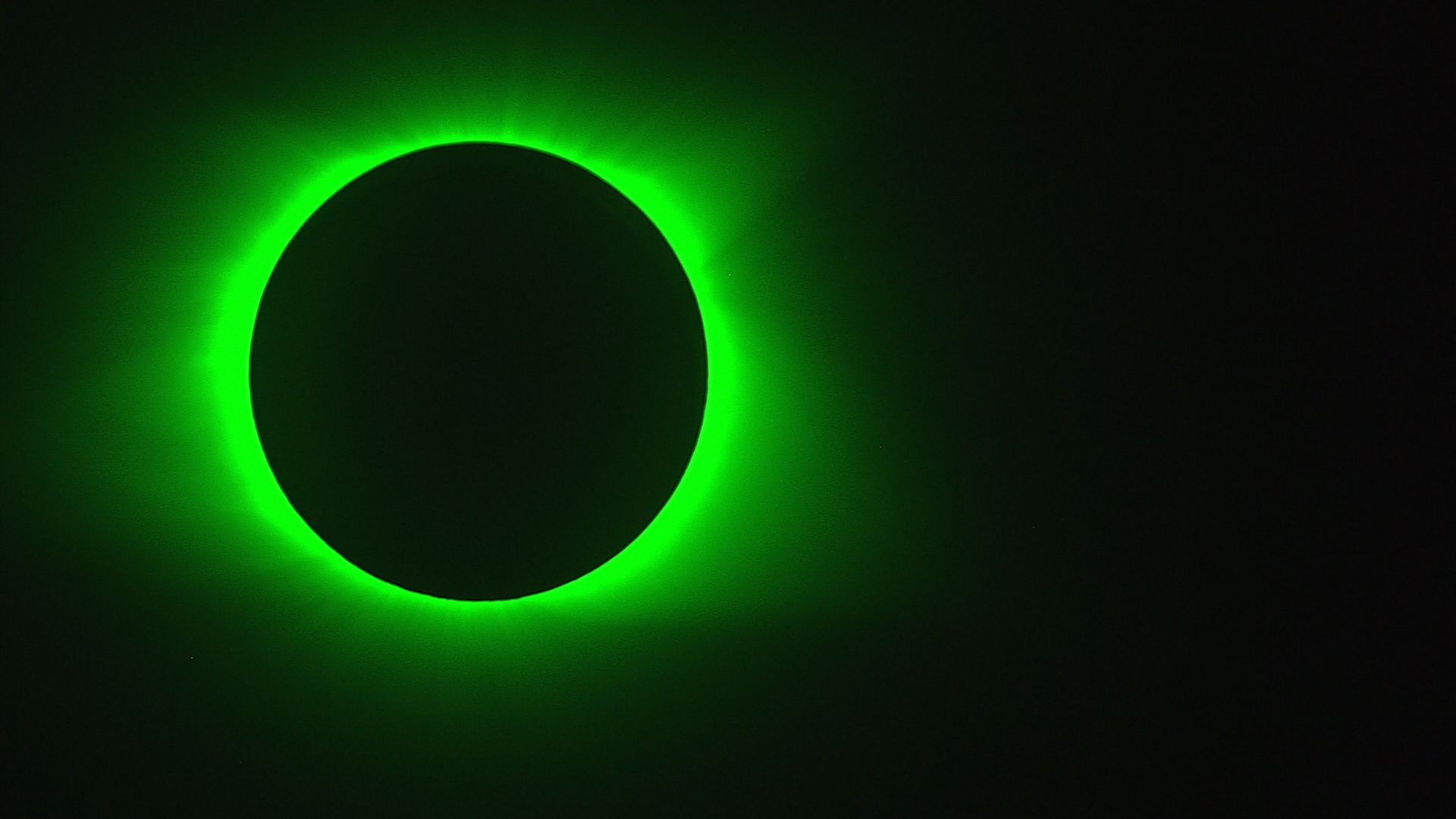}
    \caption{Raw visible-light ($533.9\pm4.75$~nm) image from the HD broadcast television camera on the leading (eastward) WB-57. The image data are polychromatic RGB, encoded as described in the main text.}
    \label{fig:raw_vis}
\end{figure*}

DyNAMITE, as currently configured, is primarily a video system which stores its visible-light data as losslessly compressed video encoded in 10-bit Y$^\prime$CbCr 4:2:2 format (SMPTE292M). This format uses a brightness (luma: Y) channel value for every pixel in the image, with two color channels (chroma: Cb and Cr) that are subsampled at half the resolution of the luma channel (essentially binned 2$\times$2). The video is then encoded in blocks of 4$\times$4 pixels. Because of the green filter, our observations are essentially monochrome, so we extract only the 10-bit luma information and discard the chroma channels. We store each frame of the video as a separate TIFF image encoded using 16 bits per pixel to preserve the complete video dynamic range. The raw, full-color video data stream is $\sim$40~GB from each aircraft over the duration of totality, yielding $\sim$60~GB total of extracted monochromatic TIFFs. Figure~\ref{fig:raw_vis} shows a single frame of the raw visible-light video during the eclipse.

The HD broadcast camera has three CCDs that internally output 3$\times$14-bit three-color (RGB) data, in which the linear CCD response has been pre-processed within the camera to compress the dynamic range. The signal is then encoded over 10 bits, which requires an additional resampling and dynamic range adjustment. The practical implication of these multiple processing steps is that the original linear CCD signal is not fully recoverable from the encoded video. To address this limitation and restore the observations to as near to linearity as possible, we characterized the camera's performance in the lab before flight using a linearly variable light source and multiple exposure times. This characterization showed how the video-encoded signal varied across its range from zero to complete saturation, and allowed us to apply a correction to restore the signal to quasi-linearity, particularly near the top of the camera's dynamic range where the nonlinear response was most pronounced.

\begin{figure*}
    \centering
    \includegraphics[width=0.95\textwidth]{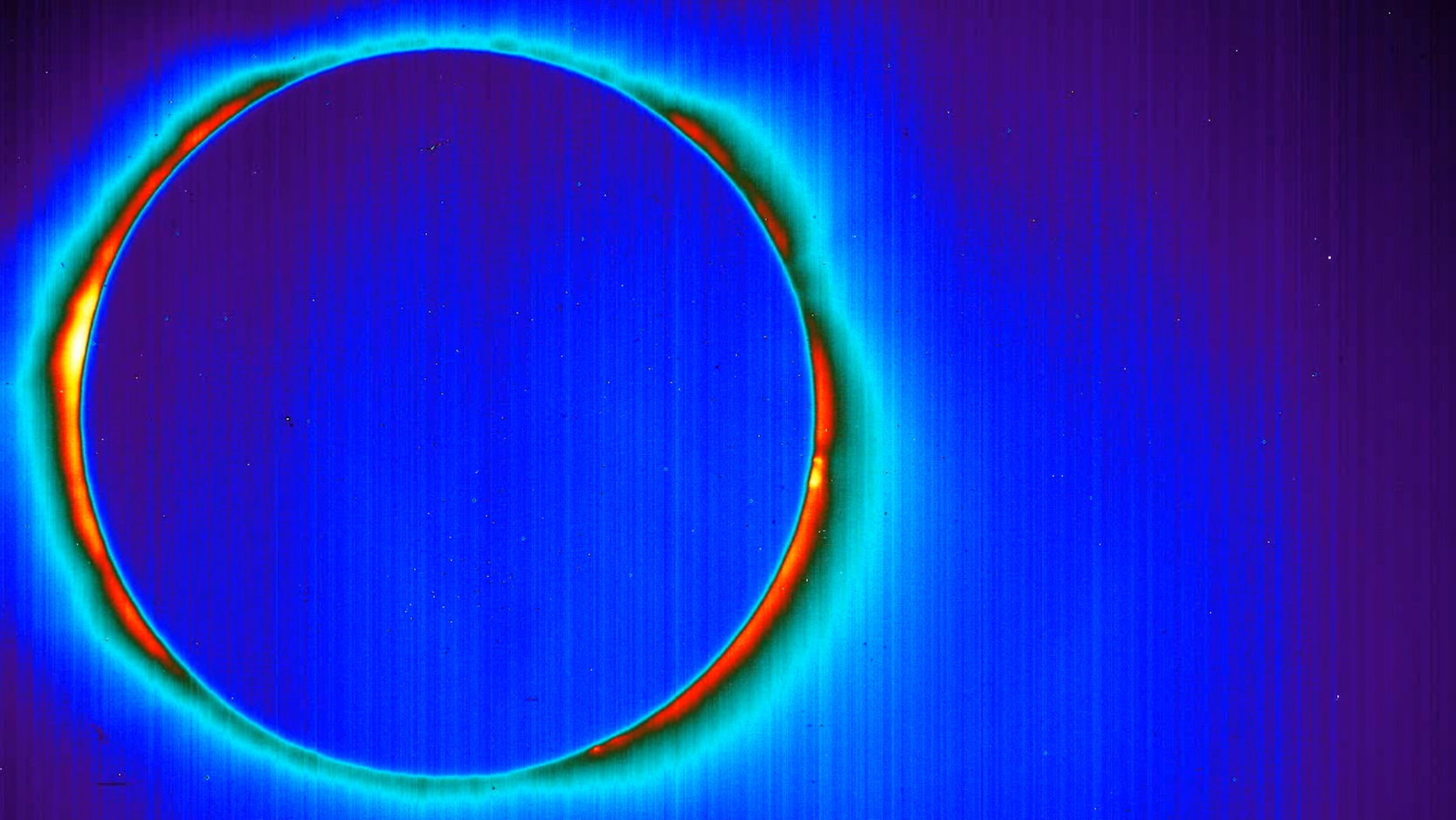}
    \caption{Raw 3--5~$\mu$m image from the MWIR camera on the trailing (westward) WB-57. The image data are monochrome, but are displayed here using a rainbow color table (black/blue is dimmest, yellow/white is brightest; the color scale is essentially arbitrary in these unprocessed images). Fixed-pattern noise is evident throughout the image but can be removed in processing. A bright prominence is visible on the west limb, and an active region on the east limb.}
    \label{fig:raw_ir}
\end{figure*}

The main calibration steps applied to the visible-light data, after linearity corrections, were to remove camera bias and dark current. We attempted to obtain flat-field observations during the flights, but found that instrumental noise and the presence of artifacts in these images limited their utility for calibration. There is some low-level, low-frequency variation across the full field of view due to the optical vignetting function of the lens, but this variation is negligible on short scales. At high frequencies the variation in the flat field is completely dominated by uncorrectable detector noise that introduced non-negligible pixel-to-pixel variations in individual frames. Consequently, we found that the processed observations had overall lower noise when we did not apply any flat-field. The only calibration step currently applied is background subtraction using dark images obtained immediately after totality. \edit1{These calibration challenges point to fundamental limitations of the existing visible light imaging system that may necessitate the addition of a new, scientific-quality camera for future flights, which we discuss in detail in Sec~\ref{sec:disc}.}

\authorcomment1{This paragraph moved from above.} The MWIR camera is designed as a scientific instrument: it has a linear response to incident flux, yields direct digital output of individual images at full 16-bit depth, and makes use of lossless encoding and embedded timestamps (although not GPS coordinates) within each image. This allows the IR data to be analyzed in a straightforward manner. The native data storage format for each image can be exported to FITS and analyzed with standard tools \edit1{and image data calibration techniques such as subtraction of background (e.g., dark current, thermal emission from the sky and enclosure, etc.) and flat-fielding}. Figure~\ref{fig:raw_ir} shows an example raw MWIR image of the eclipse; \edit1{detailed discussion of the calibration, processing, and analysis of these data will be presented in a future paper}.

\needspace{3\baselineskip}
\subsection{Pointing and Image Co-Alignment} \label{sec:pointing}
\begin{figure*}
    \centering
    \includegraphics[width=0.95\textwidth]{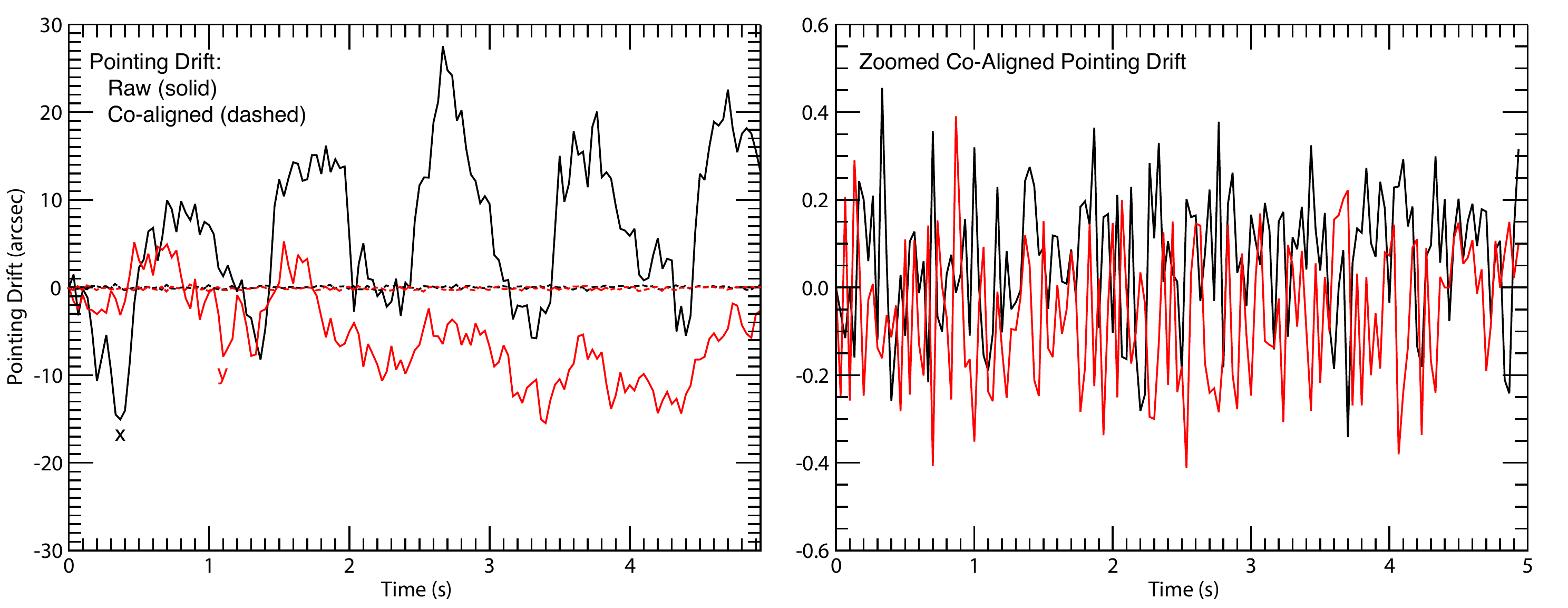}
    \caption{(Left) Measured x-y (plane-of-sky) jitter for Sirius during pre-eclipse calibration observations. The difference in x and y magnitudes of the raw jitter (solid) is likely due to the camera orientation with respect to the airplane heading and is not intrinsic to the pointing system. The inherent error in the corrected pointing (dashed) is $\sim$0.3{\arcsec} ($\sim$0.1~pixels) r.m.s. (Right) Zoomed view of corrected jitter. The residual error of $\sim$0.1~pixels r.m.s.~is at the limit of position measurement sensitivity.}
    \label{fig:pointing}
\end{figure*}

Instrument pointing and roll are not entirely stable throughout the observations for a variety of reasons, but predominantly due to in-flight turbulence causing small changes in pointing and camera orientation (due to the two-axis system) with respect to targets on the sky. There is also a small change in camera orientation with respect to the Sun introduced by the changing solar position and aircraft viewpoint over the duration of the observation. We used in-flight stellar calibration observations (specifically of Sirius) to quantify the x-y (plane-of-sky) jitter induced by these effects. For each image frame, we measured the position of the star and determined the deviation from a reference frame. We did this using multiple methods, including Gaussian fits and cross-correlation, to validate that no one method introduced significant error or bias; all methods yielded consistent results within uncertainties. To quantify the intrinsic limit of coalignment accuracy, we used these measurements to shift each image to a common reference frame and measured the residual deviations using the same methods. Figure~\ref{fig:pointing} shows the measured drift of the Sirius centroid before and after co-alignment; the zoomed view in the right panel highlights that our coalignment is within a tenth of a pixel, at the sensitivity limit for such measurements.

Our eclipse images did not include bright background stars that could be used for straightforward cross-correlative co-alignment as above. Consequently, co-registering the solar images to a common reference frame required a different, and iterative, scheme. First, we exploit the high contrast of the lunar limb as a fiducial reference to compensate for the primary jitter induced by the plane's motion. We then compensate for instrument roll (on both long and short timescales) by cross-correlating neighboring images and rotating appropriately. We further compensate for the motion of the Moon with respect to the Sun during the eclipse by shifting each image using an offset based on the \textit{a priori} known speed of the Moon's passage across the sky. To minimize error due to the multiple interpolation steps applied to achieve sub-pixel alignment of the data, we combine all the iteratively-derived corrections necessary and re-apply them to the uncorrected data in a single step at the end of the process. \edit1{Frame-to-frame jitter is largely correctable, but intra-frame jitter is a more pernicious problem because it can lead to degradation in image quality. The frame-to-frame RMS pointing error is about 1.5~arcsec --- half a pixel --- and a small enough amount of movement that most frames show no evidence of motion blur. However, the image quality in a few ($\lesssim$1\%) frames are degraded by intra-frame motion even with 30~ms exposure times. The image stacking we use to control overall image noise largely eliminates this problem in our analysis, but it is a concern that could require a mitigation strategy in future investigations where image-stacking of 30~ms exposures may not be appropriate.}

Because the solar limb is rotationally symmetric and coronal features are not entirely sharp, this iterative method is somewhat less accurate than cross-correlation of one or more bright stars, but nonetheless achieves a coalignment accuracy of $\sim$1~pixel. The 100~Hz pointing and heading information recorded by AIRS is precise enough to, in principle, allow reconstruction of the jitter and roll independently from the imaging data. We are currently evaluating whether this pointing data yields a more accurate solution. Unfortunately, an in-flight equipment failure on the trailing (westward) plane resulted in the loss of engineering (housekeeping) data that relates GPS timestamps to aircraft position, attitude, and AIRS pointing information. The iterative method described above is robust against such failures since it derives the co-alignment solution from the image data alone. For future missions, the visible-light system could be used as a guide camera rather than for scientific observations, if we optimize its configuration to capture stars.

\needspace{3\baselineskip}
\subsection{Data Quality} \label{sec:dataquality}
The visible-light data unfortunately contain several \edit1{noise sources and} artifacts that cannot be straightforwardly corrected using the calibration described earlier. One such artifact is the intermittent presence of highly structured quasi-fixed-pattern noise of variable intensity. The origin of this artifact is unknown but we suspect electromagnetic interference somewhere along the signal path. (This \edit1{artifact} was previously undetected in DyNAMITE observations, probably because the instrument has typically been used to observe high-contrast, well illuminated scenes where such low-level noise would be inconsequential.) This \edit1{artifact} primarily affects a region of the corona on the west limb, and is significant at heights of $\sim$1~R$_{\mathrm{Sun}}$ above the limb; other regions of the image are unaffected and thus have been the focus of our performance and preliminary scientific analysis efforts. The appearance of this feature during some of our dark observations should provide a path for its removal from the entire observation period, but we have not yet addressed this issue in detail.

\edit1{At a lower level, we observe variable high-frequency noise, reminiscent of film grain, throughout each image frame.  This noise is also apparent in the flat-field images (Sec.~\ref{sec:data}). Because it varies from frame to frame, it cannot be removed by subtraction or flat-fielding using standard calibration techniques.  We compensate for this by binning in time and/or space, which mitigates the effect of these small, high-frequency variations, but does not completely eliminate it (see section~\ref{sec:analysis}).}

A second, but less significant, artifact is the presence of stray light and internal reflection occurring in DyNAMITE's visible-light zoom lens. The internal reflection is most prominent in the region occulted by the lunar disk, and measurements there suggest it is at roughly 1\% of the primary signal level. Since this reflection is essentially fixed in place with respect to the coronal image, it represents an effectively static background that can be ignored in, for example, studies of coronal dynamics that are a natural target for eclipse campaigns \citep[e.g.,][]{2001MNRAS.326..428W, 2002SoPh..207..241P, 2002MNRAS.336..747W}. Stray light in the image can be thought of as the contribution of the low-level wings of a very broad instrumental point-spread function. Because the corona is such an extended source, its stray light contribution acts as a slowly varying pedestal across the whole image. This serves to slightly reduce contrast and, for studies of coronal dynamics, such a fixed offset does not contribute significantly.

Because of the rapid fall-off of coronal brightness with altitude above the solar surface -- dropping by more than 3 orders of magnitude between 1 and 3 R$_\mathrm{Sun}$ (Figure~\ref{fig:bandpass}) -- very few cameras can optimally capture the full dynamic range of coronal emission in a single exposure. While the HD broadcast television camera's 14-bit internal quantization could, in principle, accommodate this wide range, the intrinsic noise in the signal coupled with the additional artifacts discussed above necessitated a compromise between achieving good statistics in the extended corona, which was our primary science target, versus minimizing saturation in the bright, innermost corona. Therefore, we instructed the SEO to optimize the exposure time to provide the highest possible signal quality for observations in the range of roughly 1.5--3~R$_\mathrm{Sun}$, which nonetheless resulted in saturation of the inner corona out to $\sim$1.15~R$_\mathrm{Sun}$ (see Figure~\ref{fig:raw_vis}). However, since the CCD does not bloom, this saturation does not materially affect photometric analysis in the rest of the image. Because the SEO had to manually gauge the optimal exposure time while eclipse totality was in progress, each airplane's observations include a few changes in shutter speed near the beginning and/or end of the observing period, resulting in instantaneous discontinuities in the signal brightness that must be corrected during the linearization step of our data preparation. The vast majority of the eclipse observations used a $\sim$30~ms exposure time.

\needspace{3\baselineskip}
\section{Preliminary Scientific Analysis} \label{sec:analysis}

Although this was primarily an engineering mission, we had three main scientific targets for these data: searching for the presence of oscillations that could be linked to coronal heating \citep[e.g.,][]{2007Sci...317.1192T}; searching for flows that could be related to processes, such as nanoflares, that can both reshape and heat the corona \citep[e.g.,][]{2006AGUFMSH31B..07D}; and characterizing the large-scale structure of the corona to help isolate key targets for analysis and future observations. Just as planning and executing the mission as a complete scientific program provided critical insight into the systems engineering, mission design, decision-making, and operations required for success, carrying out robust, albeit preliminary, scientific data analysis allows us to quantify the performance and limitations of our data and methods in a realistic context. The results of this analysis will drive scientific priorities and corresponding instrument upgrades for future WB-57 eclipse expeditions.

Below, we discuss our preliminary analysis and results, and, more importantly, the limitations of the existing AIRS/DyNAMITE instrumention for addressing the three science targets above.

\needspace{3\baselineskip}
\subsection{Oscillations} \label{subsec:oscillations}
\begin{figure*}
    \centering
    \includegraphics[width=1\textwidth]{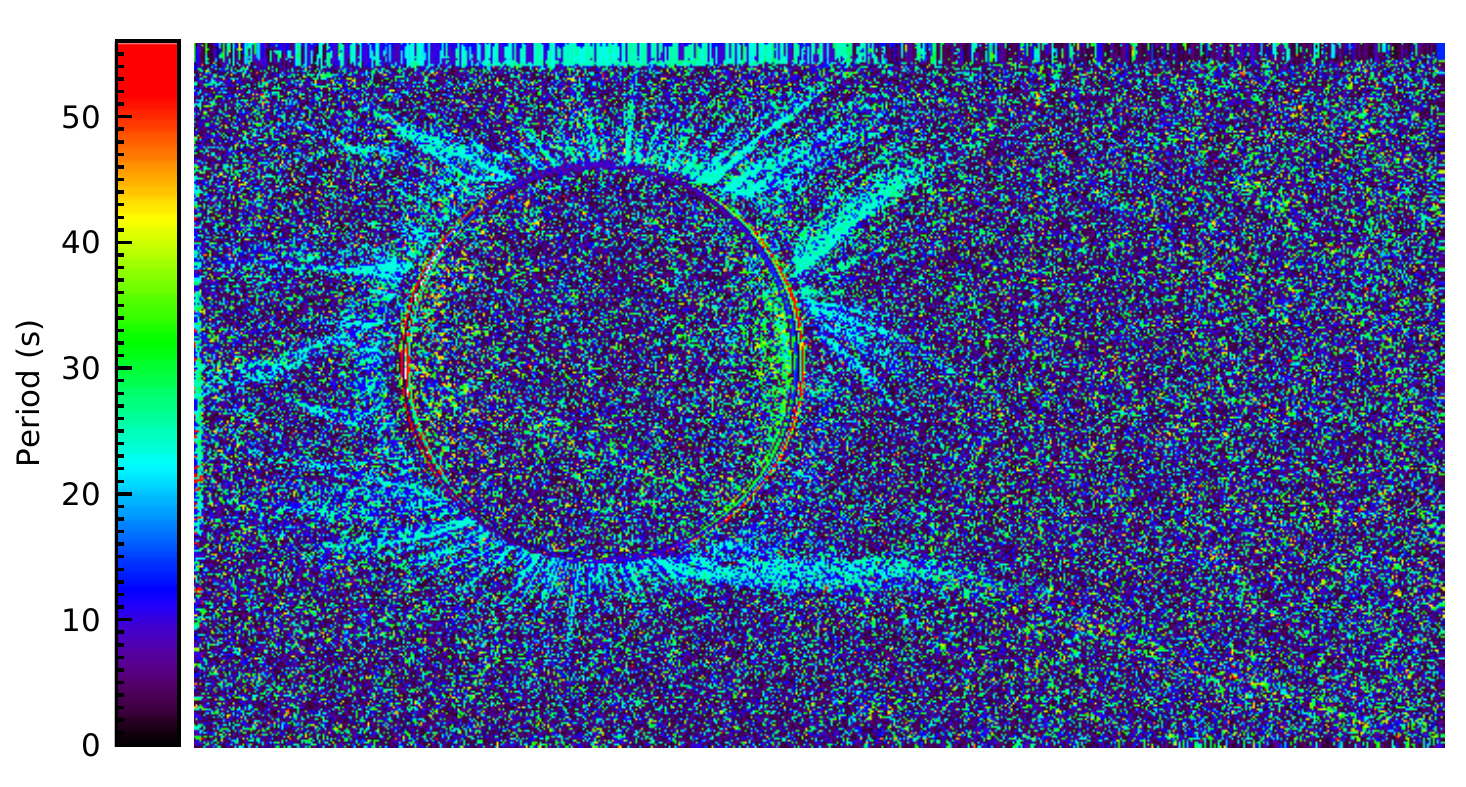}
    \caption{Period of the most prominent peak in the wavelet transform for each pixel in a three-minute subset of the green-band data from one of the planes. The cyan color that appears in the neighborhood of inner coronal structures denotes a period of $\sim$25~s.}
    \label{fig:peak_period}
\end{figure*}

The energy released by the dissipation of waves in the corona could be responsible for heating the corona to millions of degrees. Such oscillations can be observed in a variety of different modes and frequency domains; for a detailed discussion of MHD wave modes in the corona, see \citet[][chapters 7--8]{2005psci.book.....A}. High-frequency oscillations are most easily observed using ground-based instrumentation, both during eclipses and using coronagraphs, and there are reports of detections of such oscillations using both approaches \citep[see, e.g.,][]{2002SoPh..207..241P, 2007Sci...317.1192T}. One of the best examples of detections of waves during an eclipse was reported by \citet{2001MNRAS.326..428W}, who found 6-s oscillations in active region loops using Fe~\textsc{xiv} 530.3~nm observations during the 2001 eclipse. \citet{2002MNRAS.336..747W} subsequently proved that these oscillations in fact propagate along coronal loops. However, none of these prior measurements has established definitively that these waves carry sufficient energy to heat the corona to observed temperatures, and as a result, oscillations remain a natural and compelling target for eclipse studies. In particular, improvements in sensitivity, seeing, cadence, and observing duration would provide significantly improved constraints to this question.

We employed the wavelet analysis tools developed by \citet{1998BAMS...79...61T} to test for the presence of oscillatory signals in every pixel of our image sequence from one of the aircraft. To help reduce the effect of low-level noise in the data, and the impact of residual frame-to-frame misalignments, we rebinned the data $3\times3\times30$ ($x{\times}y{\times}t$), resulting in a data cube with $\sim$9{\arcsec} pixels and 1~s cadence, before taking the wavelet transform. This adjustment likely reduces our sensitivity to transverse oscillations, which are expected have an amplitude of at most 2200~km \citep{2005psci.book.....A}, corresponding to about one pixel, but should improve our sensitivity to longitudinal oscillations, which we would expect to cause coherent brightness changes over several pixels in a small neighborhood.

Our analysis shows that the wavelet transform of relatively bright regions in the observations are uniformly dominated by an oscillation with a period of about 25~s. Figure~\ref{fig:peak_period} shows a plot of the period of the strongest peak in the wavelet transform for each pixel in our dataset, which reveals an oscillation of approximately the same period that is ubiquitous throughout our observation. It is unlikely that oscillations of solar origin would be so uniform throughout the corona, which suggests that this feature is likely an artifact of \edit1{the limitations of our} calibration or pointing correction. While atmospheric turbulence around the aircraft could be responsible for a spurious oscillatory signal, we do not see any evidence of spatial distortions in the image field that would likely be characteristic of such turbulence, and the subsonic speed of the aircraft eliminates the possibility of extreme turbulence associated with a sonic shock, as has been observed by other missions \citep{1975LAstr..89..149K}. Furthermore, an oscillation with roughly this period appears to be associated with the uncorrected fixed-pattern artifact described in Section~\ref{sec:data}, in pixels where we obviously would not expect to detect oscillations with real physical origins. Thus, we conclude that this dominant oscillatory signal is almost certainly an artifact, which we aim to remove with improved data processing in the future.

In general, this \edit1{artifact-induced} oscillation is associated with fluctuations of $\sim$1--2\% of the total signal in each pixels. Thus, \edit1{the limitations of data quality inherent to the existing instrumentation and calibration place a lower limit of about 2\% on our sensitivity to oscillations of solar origin. This} is roughly the peak amplitude we would expect \edit1{from} prior studies; \edit1{however, the semi-broad filter response of the current setup also includes K-corona from overlying material in the line of sight that would likely reduce the observed amplitude even if data quality were improved}. Further improvements in calibration, which are presently underway, could facilitate the detection of weaker fluctuations that might be associated with a variety of oscillations. However, for future studies, a much more significant improvement could be realized by upgrading the visible light instrumentation, particularly replacing the broadcast-quality camera with a scientific camera, and replacing the optics to reduce internal reflection and accommodate a narrowband green-line filter \edit1{that would eliminate much of the overlying K-corona contribution and improve sensitivity to oscillations}. 

It is worth noting that some studies have suggested that waves with a local density amplitude above 10\% should be present at all times rather uniformly throughout the corona \citep{2007Sci...317.1192T}. These oscillations have much longer periods (roughly 5~minutes) and thus could only be detected in our data after we unify the observation sets from both airplanes. Unfortunately, this task has been complicated by the loss of the engineering data from the trailing airplane, although a bootstrap method to unify the two data sets is in progress (see Section~\ref{sec:disc}). 

\needspace{3\baselineskip}
\subsection{Flows in the Corona} \label{subsec:motion}

\begin{figure*}
    \centering
    \includegraphics[width=0.95\textwidth]{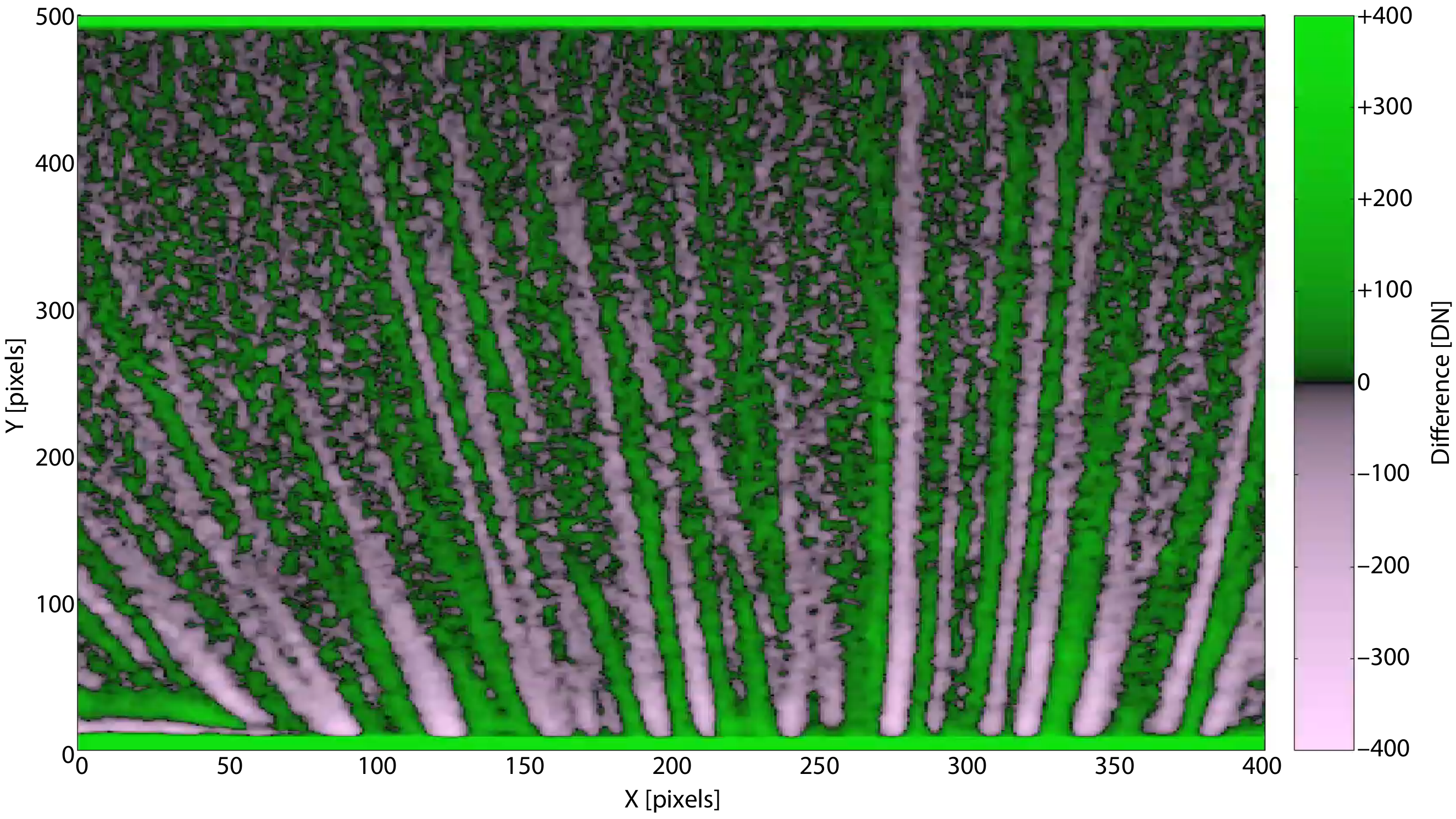}
    \caption{Mean-subtracted green-band image of the polar plumes near the north pole during the eclipse. This image is a 1-s (30-frame) average with the mean of $\sim$3 minutes of co-aligned observations removed to enhance the visibility of dynamic features.  Plumes are still clearly visible in individual frames because residual drift leads to smoothing in the three-minute mean image that is subtracted.  The color table is optimized for differences of $\pm$50 DN and up, against a background of a few thousand DN, or $\sim$1--2\% of the total brightness in the field of view.  We used time series of images like this one to search for high frequency waves in the corona. (A movie is available in the online supplementary materials.)}
    \label{fig:plumes}
\end{figure*}

Plumes have long been known to host propagating periodic (or quasi-periodic) motions that are believed to be associated with nascent solar wind \citep{McIntosh2012}. \citet{Liu2015} observed such motions using the Atmospheric Imaging Assembly \citep[AIA;][]{Lemen2012} on the \textit{Solar Dynamics Observatory} \citep[SDO;][]{Pesnell2012} and found them to propagate at around 120~km~s$^{-1}$. Such features could be expected to propagate $\sim$10~pixels over the course of a few minutes at our $\sim$3{\arcsec}/pixel platescale, and thus should be easily detectable in our native-resolution observations if they are present. Plumes therefore present a promising location to begin a search for motions in the corona in our own data. Additionally, polar plumes provide a high-contrast environment with minimal line-of-sight contamination, where small scale motions can be detected without interference or confusion resulting from other features in the foreground or background. 

We searched for motions by first preparing a de-noised version of our data set using the noise-gating technique described by \citet{DeForest2017} to reduce the effects of photon-counting and detector-induced noise as discussed in Section~\ref{sec:data}. We then attempted to isolate motions by removing a long-term average from every frame in the movie, leaving only dynamic features in each frame. Figure~\ref{fig:plumes} and the associated animation shows the resulting processed data for a small section of the plumes in the north polar region.

The image features remaining after this processing include slow-moving, nearly-radial striae and small (few-pixel) blobs. Typical quasi-stationary features have amplitudes of $\pm$100--200~DN against a coronal background of a few thousand~DN. Small blob-like fluctuations are present with amplitudes of order $\pm$20~DN, but are not readily separable from residual noise. Time/space Fourier analysis did not reveal clear asymmetry in propagation direction, which could indicate symmetric motion in the corona but is also consistent with the null hypothesis that the blobs are residual camera noise and/or the artifacts resulting from residual errors in image co-alignment. Based on the observed amplitudes of the striae, which are well detected, and the blobs, which are not, we can place an upper limit on small transient phenomena in this time/space size range of 0.3--0.5\% of the overall brightness. As in Section~\ref{subsec:oscillations}, a future analysis using improved calibration and image co-alignment will be required to better suppress artifacts to allow us to isolate any possible dynamics, but far more significant gains can be made on future flights using upgraded instrumentation.

\needspace{1\baselineskip}
\subsection{Coronal Structure} \label{subsec:structure}

Because the corona spans such a wide dynamic range of brightness even within a few R$_{\mathrm{Sun}}$ above the limb, studies of large-scale structures require special image processing to reveal coherent features with sufficient contrast to track them throughout the corona \citep[e.g.,][]{2006CoSka..36..131D}. Likewise, achieving adequate signal-to-noise to observe faint features high in the corona while simultaneously avoiding saturation in the low corona requires ``high dynamic range'' (HDR) imaging through the use multi-exposure composites \citep[e.g.,][Section~5.3]{2019A&A...622A.101S} or stacking of multiple images with relatively short exposure times. The current AIRS/DyNAMITE camera does not natively support multi-exposure HDR -- the exposure time is effectively fixed unless changed manually -- and we thus used the latter method. As discussed in Section~\ref{sec:data}, the necessary optimization compromises still resulted in a saturated innermost corona, but provided the required sensitivity at higher altitudes and provided the ability to correct for frame-to-frame jitter in a uniform manner.

\begin{figure*}
    \centering
    \includegraphics[width=0.95\textwidth]{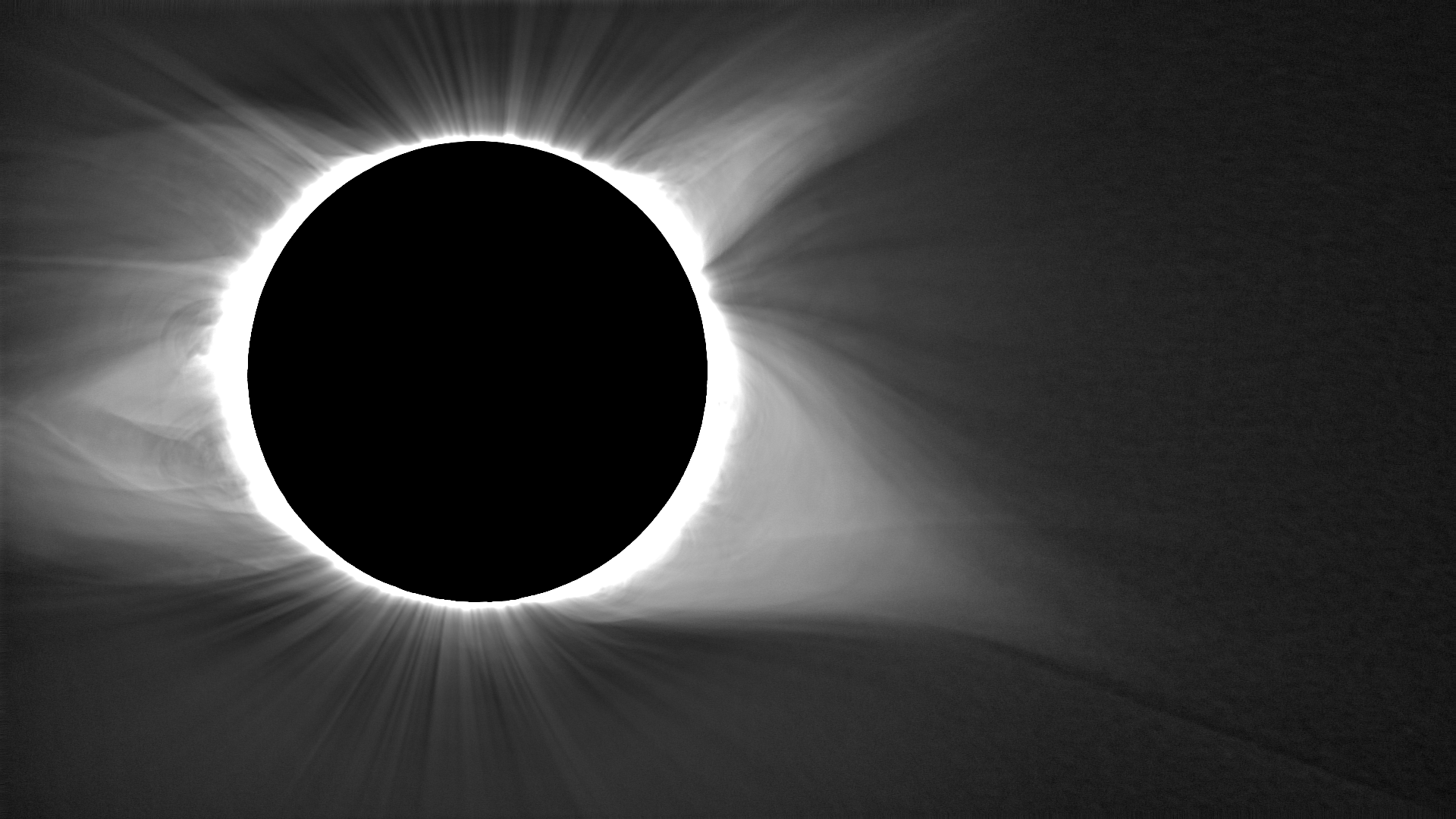}
    \caption{Fully processed image of the corona generated by stacking many co-aligned, calibrated images and applying a radial filter and multi-scale unsharp mask to bring out details at higher altitudes. Note that we have retained the original image orientation with solar north $\sim$6$^{\circ}$ clockwise from top.}
    \label{fig:filtered_eclipse}
\end{figure*}

We generated a high-quality image of the corona by median-stacking many calibrated, co-aligned images for an effective exposure of $\sim$120~s. We then rescaled the composite image using an azimuthally uniform filter derived from the median falloff of the coronal brightness as a function of height, and applied a multiscale unsharp mask to enhance feature contrast (Figure~\ref{fig:filtered_eclipse}). Although the inner corona is saturated out to as far as $\sim$1.15~R$_\mathrm{Sun}$ in this image, we are able to track coronal structures to heights of $\gtrsim$2~R$_\mathrm{Sun}$ in the east and $\gtrsim$4~R$_\mathrm{Sun}$ in the west.

\begin{figure*}
    \centering
    \includegraphics[width=0.32\textwidth]{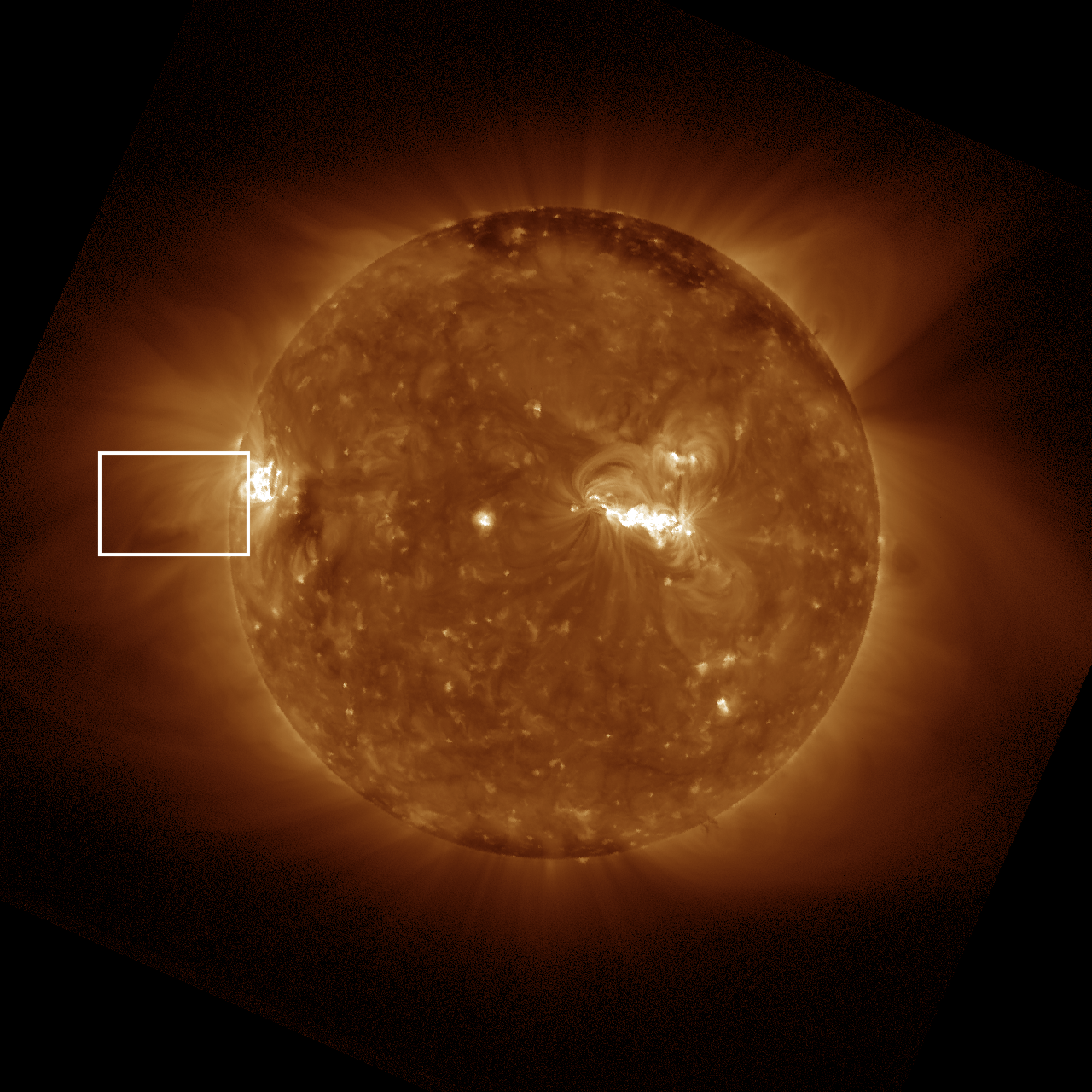}
    \includegraphics[width=0.32\textwidth]{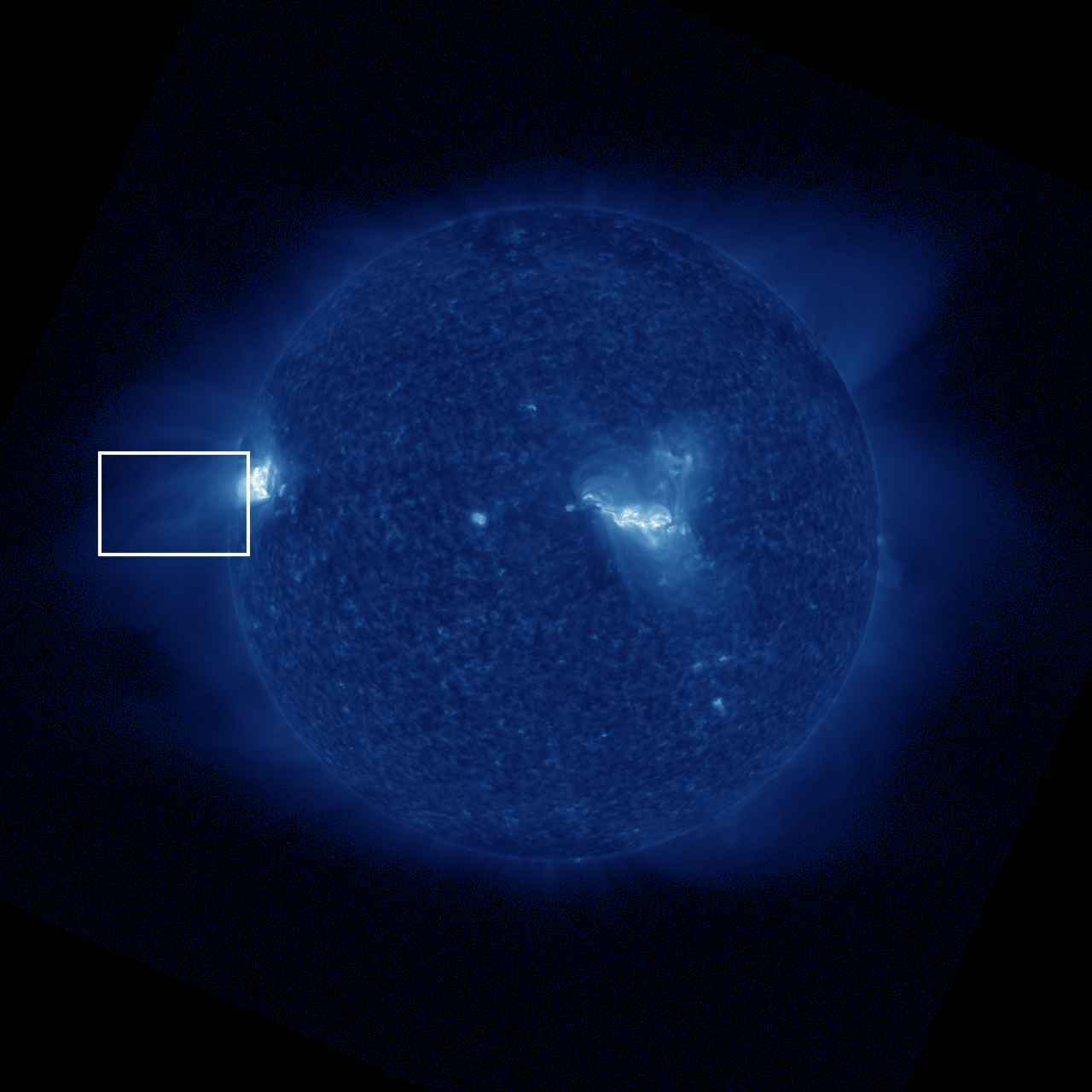}
    \includegraphics[width=0.32\textwidth]{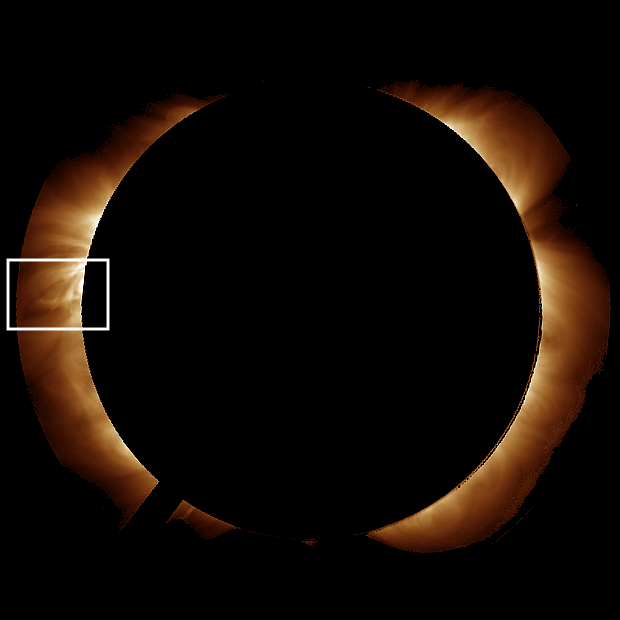}
    \caption{The corona at the time of the eclipse from \textit{GOES-16}/SUVI, in the 195~{\AA} (\textit{left}; Fe~\textsc{xi--xii}, $\sim$1.5~MK) and 284~{\AA} (\textit{center}; Fe~\textsc{xv}, $\sim$2~MK, plus He~\textsc{ii} 304~{\AA}) passbands, and from CoMP, in the 10747~{\AA} (\textit{right}; Fe~\textsc{xiii}, $\sim$1.8~MK) passband. Note that the image platescales are not identical and the images have been rotated from their native orientations to match the WB-57 eclipse image frame (Figure~\ref{fig:filtered_eclipse}), with solar north $\sim$6$^{\circ}$ clockwise from the top of the image. \edit1{The white box highlights a void in the SUVI 284~{\AA} image where bright loops are present in the other panels and in our WB-57 green-band image.}}
    \label{fig:suvi_eclipse}
\end{figure*}

The large scale structure of the corona at the time of the eclipse was dominated by three features: a large streamer/active region complex on the east limb, an apparent pseudostreamer in the northwest, and a large helmet streamer in the southwest, the latter two of which were our primary observation targets. The Solar Ultraviolet Imager \citep[SUVI; see][for a brief instrument summary]{Seaton2018} on the \textit{GOES-16} spacecraft provides wide-FOV observations in multiple extreme ultraviolet passbands that shed light on the physical properties (e.g., temperature) of the structures we observe (Figure~\ref{fig:suvi_eclipse}). A comparison of SUVI images to our WB-57 eclipse observations reveals how the large-scale corona is connected to the surface features that drive its complexity. Two SUVI passbands are dominated by emission lines from Fe ionization states closest to the Fe~\textsc{xiv} that contributes to our green-band eclipse image. In particular, SUVI's 195~{\AA} band includes a variety of Fe~\textsc{xi--xii} lines with temperatures near 1.5~MK, while the broad 284~{\AA} band is predominately Fe~\textsc{xv} at about 2~MK, with a small contribution from the chromospheric He~\textsc{ii} line around 304~{\AA}.

The Coronal Multichannel Polarimeter \citep[CoMP;][]{Tomczyk2008} at the Mauna Loa Solar Observatory is a near-infrared coronagraph, imaging the corona from $\sim$1.05--1.4~R$_\mathrm{Sun}$ in two Fe~\textsc{xiii} lines near 10747~{\AA}. CoMP observations reveal the same prominent loops on the east limb, related to Active Region (AR)~12672, as are clearly visible in both the SUVI eclipse images. A void in the southern half of this structure, visible in SUVI 284~{\AA} \edit1{(see white box in Figure~\ref{fig:suvi_eclipse})}, is not obvious in the WB-57 green-band, CoMP 10747~{\AA}, or SUVI 195~{\AA}, suggesting the void is very likely the result of a strong temperature gradient in the corona and is populated by material too cool to appear in Fe~\textsc{xv} 284~{\AA}. \edit1{This interpretation is supported by \citet{2020ApJ...888..100B}, whose narrowband eclipse images in Fe~\textsc{xi} and Fe~\textsc{xiv} show the same temperature-sensitive effect. This also highlights that our WB-57 green-band image, which shows the loops but not the void, has a significant contribution from the temperature-insensitive K-corona.} This temperature gradient might be related to active, anisotropic heating that could make this region an interesting target for both dynamics and oscillations studies following calibration and pointing improvements. Intriguingly, the void appears to be co-spatial with an \edit1{infrared} intensity enhancement on the east limb, \edit1{the base of which is visible} in the unprocessed \edit1{MWIR} image (Figure~\ref{fig:raw_ir}), that we will explore in a future paper.

Although our WB-57 green-band observations do reveal fine details in the corona, the airborne platform conveys little to no clear advantage for ultra-high-resolution studies of static coronal structures \citep[e.g.,][Figure~1A]{2018NatAs...2..913M} compared to optimized ground-based instrumentation, in contrast to studies of dynamics where both duration and seeing are the critical factors. Without the constraints imposed by the platform, ground-based equipment can employ, for example, much larger apertures and/or much narrower-passband filters. Likewise, observing techniques such as lucky imaging \citep{2016ASSL..439....1B} can largely mitigate the degradation caused by seeing effects for static images. Furthermore, the extra time in totality afforded by the mobile airborne platform is of limited value for static imaging compared to these ground-based advantages. Upgraded instrumentation for AIRS/DyNAMITE can certainly improve the quality of our airborne static imaging results in future missions; however, it is clear that the biggest gains in performance for this platform will be realized in studies of coronal dynamics and/or by optimizing for wavelengths not accessible from the ground, such as mid-wave infrared.

\needspace{3\baselineskip}
\section{Discussion} \label{sec:disc}

The NASA WB-57F with AIRS/DyNAMITE presents a viable platform for solar astronomy, benefiting from mobility and minimization of atmospheric effects -- including absorption, emission, and turbulence due to its high-altitude flight -- with a stable two-axis pointed optical bench. We successfully tested and characterized the capability and limitations of this platform using the existing DyNAMITE visible and infrared instrument. For this pathfinding engineering mission, we leveraged the existing instrumentation to maximize strictly limited resources, even though this instrumentation is not optimized for scientific measurements. Future flights with enhancements to DyNAMITE to improve its response, or entirely new instrumentation, will make solar observations at reasonable cost that are currently infeasible or prohibitively expensive using other platforms. Such upgrades could also enable this platform for general solar observations outside of eclipses.


To best understand the performance of the existing AIRS/DyNAMITE platform in an appropriate scientific context, we adopted three scientific objectives targeted to the advantages offered by this platform. Treating the engineering-quality data with the full rigor of a scientific mission provided deep insight into the platform performance and resulting data quality to drive decisions about upgrades and revised objectives for future missions. This preliminary analysis yielded upper limits on the amplitude of dynamic phenomena, including both flows and oscillations in the corona.

The idiosyncracies of the repurposed DyNAMITE instrument, largely resulting from the nonscientific nature of the visible-light camera and optics, limited our sensitivity to both dynamics and fine-scale structure. Some fundamental constraints can only be resolved by instrumental modifications for future flights. However, ongoing calibration efforts will identify camera parameters and operational techniques that can improve data quality for future missions even with the existing camera/optics system. 

In contrast, AIRS/DyNAMITE's existing MWIR camera is a scientific instrument, not subject to the same limitations as the visible light system, and has shown significant potential for scientific exploitation even in its current configuration. Its 3--5~$\mu$m wavelength range has been poorly explored for coronal science, in large part due to the significant difficulty in observing this range from ground-based facilities. As a result, we had little indication prior to flight of what the MWIR observations would reveal, nor any way to predict what coronal features -- if any -- would be prominent in the data. Indeed, there are a number of relatively strong emission lines in the 3--5~$\mu$m range, sensitive to a broad range of coronal temperatures \citep{Judge1998}, such as Si~\textsc{ix} ($\sim$1.2~MK) at $\sim$3.9~$\mu$m \citep{Judge2002}, as well as to chromospheric temperatures, such as H~\textsc{i} and He~\textsc{i} ($\lesssim$35~kK) between 4 and 5~$\mu$m \citep[][]{2015A&A...582A..56D}. A number of the coronal lines are also sensitive to the magnetic field \citep[][]{Judge1998}, so MWIR observations can open brand new discovery space in coronal and chromospheric measurements. In fact, a rudimentary analysis of the IR observations during early totality has revealed intriguing large-scale structure in the east-limb corona as well as \edit1{a prominence} on the west limb. \edit1{There has been very little discussion of the nature of prominences in this MWIR range in the literature. Combining our WB-57 data with observations by the AIR-Spec spectrometer \citep{2018ApJ...856L..29S} and/or a ground-based narrowband 3.9~$\mu$m campaign could allow us to quantify the contributions from line and continuum emission, and scattering of photospheric light.} Calibration of the IR data and further analysis of both structure and dynamics is ongoing and will be the subject of a future paper.


One critical success in mission planning and implementation was the coordination of two separate WB-57 aircraft to make precisely timed observations of totality to nearly triple the contiguous observing window well beyond what could be achieved from a single stationary vantage. Work still remains to integrate the two separate sets of observations from the two respective planes into one continuous observation of totality. Despite the loss of engineering data from the trailing airplane, the high-quality imaging data is intact. We have identified a strategy by which we can bootstrap the required timing information from concurrent MWIR images, where timestamps are embedded, coupled with ground-recording of the aircraft transponder data to reconstruct exact position, pointing, and time for each frame in the visible dataset. The MWIR dataset, in contrast, is more straightforward to integrate since it includes embedded GPS timestamps. This work is ongoing and will inform both planning and potential platform improvements for future tandem-flight observing campaigns.

An imminent window for additional eclipse studies well-suited to the WB-57 platform is upcoming on 2020 December 14 over South America. Given our successful demonstration of the viability of this platform for this type of observation, and the considerable engineering and operational knowledge obtained from this first mission, this upcoming total eclipse provides a valuable opportunity for reflight that implements lessons learned\deleted{ and potential instrumentation upgrades}. \edit1{Our evaluation of the IR data quality and prospects for scientific exploitation suggests that} the 3--5~$\mu$m wavelength range \edit1{offers the largest potential for additional discovery even without}\deleted{could become a new area of focus that may not require} major modifications to the existing DyNAMITE MWIR system.

The next total solar eclipse over the U.S., on 2024 April 8, provides even greater potential to exploit this new platform for eclipse observations. Using the experience gained from our 2017 mission and the upcoming 2020 opportunity, and the results of our scientific analyses of these data, we can identify the specific measurements and associated requirements needed to address our scientific objectives about coronal dynamics and structure. With approximately 4~years (as of this writing) before this event, we have sufficient time remaining to plan and implement optimized operational procedures and potential instrumentation upgrades or modifications to provide a highly focused mission targeted precisely at obtaining the required measurements to resolve these long-standing questions.

\edit1{For the visible-light system, obvious instrumentation upgrades would include replacing the current camera and lens with a scientific-grade CMOS-APS camera and simplified mirror-based optics. Low-cost, off-the-shelf cameras and small telescopes are already widely used for similar studies, e.g., for observations of stellar occultations, and would be a high-value, low-risk change. A filter wheel could also be integrated with the new optics. For the MWIR system, the primary upgrades would be a filter wheel or other modifications to fine-tune the passband. Exploiting the existing AIRS platform removes the need to interact directly with aircraft mechanical and electrical systems, simplifying the integration process to require only changes to the optical bench. The most significant effort is likely to be software modification to integrate any new electronics into the existing AIRS user control interface operated by the SEO. Such modifications could also add support for quantitative focus and dynamic range feedback to the SEO.}

Like the 2017 eclipse, the 2024 shadow path crosses large tracts of highly-populated and easily-accessible terrain. This makes it \edit1{both} straightforward \edit1{and valuable} to coordinate with multiple deployable ground-based instruments \citep[e.g., Citizen CATE;][]{Penn2017} to obtain supplementary and contextual observations. The WB-57 fuselage and wings also provide multiple accommodation options for other instrumentation with various view-angles, \edit1{providing further opportunity for coordination with other complementary scientific programs}, including in adjacent fields such as aeronomy or atmospheric physics.

We have focused our efforts with the WB-57F on eclipse observations partly because they could be demonstrated using the existing instrumentation with little or no modifications, and partly because these events are natural targets to highlight the capabilities of a mobile, high-altitude observatory. The flexibility of the AIRS pointing platform enables the possibility of a customizable optical bench suitable for a wide range of astronomical observations. With its relatively modest aperture, the WB-57 is particularly well suited for observations of daytime targets (in addition to the Sun) not accessible by larger facilities such as SOFIA and which are difficult to observe from the ground due to prominent sky brightness and other atmospheric effects. The WB-57's high-speed satellite link enables real-time communication with mission scientists, including data downlink and potential remote instrument operation, regardless of deployment location, reducing the need for complicated travel logistics. Our 2017 total solar eclipse mission, the first of hopefully many with the WB-57s, opens the door to expanding the capabilities of NASA's Airborne Science Program and a new era in airborne astronomy.



\acknowledgments
This work was funded by NASA grant NNX17AI71G. The National Center for Atmospheric Research is sponsored by the National Science Foundation. We thank Andy Roberts and Derek Rutovic for consultation and advice; and Lika Guhathakurta, Dan Moses, and Jeff Morrill for facilitating the use of the unique WB-57 and AIRS/DyNAMITE platform that enabled this mission. We are grateful to all of the NASA WB-57 ground and air crew for their service in making this mission successful; Southwest Research Institute and Southern Research for their internal contributions above and beyond expectations; and Viasat for their generous donation of the live satellite feeds from both aircraft that enabled essential real-time feedback between scientists, engineers, and SEOs. We also thank the anonymous referee for helpful comments that improved this paper.

Data from the mission is available from the project website (\url{https://eclipse.boulder.swri.edu/}). Because of its size, raw data is currently only available through physical media. Calibrated data will be made available for download, potentially through the NASA SDAC and/or VSO, after the refined calibration and validation is complete.

\facility{WB-57 (AIRS/DyNAMITE), GOES (SUVI), HAO (CoMP)}


  \bibliographystyle{aasjournal_2019.bst}
  \bibliography{eclipse.bib}




\end{document}